\documentclass[aps,prd,nofootinbib,showpacs,superscriptaddress,twocolumn,floatfix,cite]{revtex4-1}

\usepackage{amsmath,amssymb,bm,color,fancyhdr,hyperref,lastpage,units,soul}
\usepackage{graphicx}
\graphicspath{{figures/}}

\usepackage[latin1,utf8]{inputenc}
\usepackage[english]{babel}
\usepackage{makecell}

\newcommand{\pbp}{\langle\bar{\psi}\psi\rangle}
\newcommand{\Minv}{D^{-1}}
\newcommand{\tr}{\text{tr}}

\newcommand{\chic}{\chi_\text{con}}
\newcommand{\chid}{\chi_\text{disc}}
\newcommand{\chit}{\chi_\text{tot}}

\newcommand{\fK}{f_K}

\newcommand{\mc}{m^c_q}
\newcommand{\ms}{m^\text{phys}_s}
\newcommand{\Ns}{N_\sigma}
\newcommand{\Nt}{N_\tau}
\newcommand{\Tpc}{T_{pc}}
\newcommand{\Tpcl}{T_{pc}}

\newcommand{\Hbd}{H^{1/\beta\delta}}

\newcommand{\chisq}{\chi^2/\text{dof}}

\def\lsim{\raise0.3ex\hbox{$<$\kern-0.75em\raise-1.1ex\hbox{$\sim$}}}
\def\gsim{\raise0.3ex\hbox{$>$\kern-0.75em\raise-1.1ex\hbox{$\sim$}}}

\begin{document}
\title{The Chiral Phase Transition in three-flavor QCD from Lattice QCD}

\author{Lorenzo Dini}
\affiliation{Fakult\"at f\"ur Physik, Universit\"at Bielefeld, D-33615 Bielefeld, Germany}

\author{Prasad Hegde}
\affiliation{Centre for High Energy Physics, Indian Institute of Science, Bangalore 560012, India}

\author{Frithjof Karsch}
\affiliation{Fakult\"at f\"ur Physik, Universit\"at Bielefeld, D-33615 Bielefeld, Germany}

\author{Anirban Lahiri}
\affiliation{Fakult\"at f\"ur Physik, Universit\"at Bielefeld, D-33615 Bielefeld, Germany}

\author{Christian Schmidt}
\affiliation{Fakult\"at f\"ur Physik, Universit\"at Bielefeld, D-33615 Bielefeld, Germany}

\author{Sipaz Sharma}
\affiliation{Centre for High Energy Physics, Indian Institute of Science, Bangalore 560012, India}

\begin{abstract}
We analyze the pseudo-critical behavior of three-flavor QCD using highly improved staggered quarks (HISQ) on lattices with temporal extent
$N_\tau =8$ and for quark masses corresponding to a pseudoscalar Goldstone mass in the range 
$80 ~ {\rm MeV} ~ \lsim ~ m_\pi ~ \lsim ~ 140 ~ {\rm MeV}$. Our findings are consistent with the occurrence of a second order chiral phase transition at vanishing values of the quark masses. The chiral phase transition temperature at this finite value of the lattice spacing is determined to be $T_c = 98_{-6}^{+3}~{\rm MeV}$. 
A comparison with a corresponding analysis performed in (2+1)-flavor QCD suggests that the continuum limit extrapolated chiral phase transition temperature in three-flavor QCD
will turn out to be below 90~MeV.
\end{abstract}
\date{\today}
\maketitle

\section{Introduction}
\label{sec:introduction}
The nature of the QCD chiral phase transition as a function of the number of light flavors and the value of the quark masses,
has been the subject of intense and ongoing study ever since the first work of Pisarski and 
Wilczek~\cite{Pisarski:1983ms}. 
It is now well-established that the transition is a crossover for the physical values of the light and strange quark 
masses~\cite{Aoki:2006we}. The transition is, however, expected to become second order
and first order in the two flavor and three flavor chiral limits, respectively. Consequently, a 
critical value for the strange quark mass is 
expected to exist where a tri-critical point separates the region of first and second order 
chiral phase transitions. This is sketched 
in the well-known Columbia plot for the quark mass
dependence of phase transitions in QCD
\cite{Brown:1990ev}.

Lattice studies have provided strong evidence for the expected second order nature of the chiral phase
transition in the 2-flavor chiral limit 
also for physical value of the strange quark mass.
The transition is expected to belong to the $3$-$d$, $O(4)$ universality class if the anomalous $U(1)_A$ symmetry remains sufficiently broken; otherwise, a second order phase transition belonging to the $U(2) \times U(2)$ universality class is also possible \cite{Pelissetto:2013hqa}. However, as critical exponents in both universality classes are similar a differentiation
between both possibilities will be difficult in practice \cite{Pelissetto:2013hqa}.

While the axial anomaly plays a decisive role for the symmetry breaking pattern in 2-flavor QCD, it is of less relevance in three-flavor QCD \cite{Pelissetto:2013hqa}.
Irrespective of whether or not the $U(1)_A$
symmetry is effectively restored at the 
chiral phase transition temperature, a 
renormalization group (RG) analysis \cite{Resch:2017vjs} suggests that the transition is first order, as
originally predicted in \cite{Pisarski:1983ms}.
The functional RG (FRG) analysis of the three-flavor chiral 
transition, however, suggests that a first order transition, only
occurs for rather small values of the
pion mass, $m_\pi \lsim 25$~MeV. This is in accordance with lattice QCD calculations
using staggered fermions. Although such calculations find first order transitions
in calculations with unimproved gauge and fermion actions \cite{Karsch:2001nf},
the bounds on the critical mass show a strong cut-off and discretization scheme dependence
\cite{Karsch:2003va,Varnhorst:2015lea,deForcrand:2017cgb}.  
In calculations using
improved staggered fermions no direct 
evidence has been found for a first 
order transition on lattices with temporal
extent $N_\tau=6$ for $m_\pi \gsim 80$~MeV
and a bound on the pseudoscalar Goldstone mass, above which
no first order transition exists, has been
estimated to be $m_\pi^c \simeq 50$~MeV
\cite{Bazavov:2017xul}. In calculations
with ${\cal O}(a)$ improved Wilson
fermions \cite{Kuramashi:2020meg} first order
transitions have been found at non-zero values of 
the quark masses and
the bound on the critical mass is weaker,
$m_\pi^c \lsim 110$~MeV. This bound, however,
also is consistent with a continuous transition 
in the continuum limit
\cite{Kuramashi:2020meg,Cuteri:2021ikv}.
In fact, a recent analysis of the order 
of the chiral transition as function of the 
number of flavors, performed with staggered
fermions and extrapolated to the continuum limit
\cite{Cuteri:2021ikv}, suggests that the chiral
phase transition in three-flavor QCD is second order,
which is in contrast to RG analyses. However,
as has been pointed out, such analyses are based
on a Landau-Ginsburg effective action for the 
order parameter, which is arrived at by integrating out all gauge degrees of freedom ending up with
a $\phi^4$ effective Lagrangian for the order parameter field.  
The role of gauge fluctuations, however, is subtle and may also influence the order of the chiral phase
transition \cite{Pelissetto:2017sfd,DElia:2018fjp}.
It also has been argued that a $\phi^6$ contribution to the effective Lagrangian for the order parameter may be of relevance and may allow for a second order
chiral phase transition to occur in QCD with 
number of massless flavors being larger than two
\cite{Cuteri:2021ikv}. A continuous transition
in the chiral limit of three-flavor QCD thus may not
be ruled out entirely.

In addition to the exploration of the flavor
dependence of the chiral phase transition the
determination of the chiral phase transition 
temperature is of central interest. The 
chiral phase transition temperature is expected
to drop with increasing number of flavors, $n_f$,
\cite{Braun:2006jd} and  will eventually vanish 
at a critical value of the number of flavors,
corresponding to the conformal limit of
QCD \cite{Lombardo:2014mda}, $n_f^*\sim 10$. 
In recent years, different lattice estimates for both the chiral crossover $T_{pc}$
as well the 2+1-flavor chiral transition temperature $T_c^{n_f=(2+1)}$, obtained using
different actions and with different choices of observables for setting the scale,
have converged \cite{Bazavov:2018mes, Borsanyi:2020fev, Ejiri:2009ac, Ding:2019prx}.
The crossover temperature, occurring for physical values of the two (degenerate)
light quark masses ($m_u=m_d$) and a strange quark mass, $\ms\simeq 27 m_u$, has
been determined to within 1\% to be $\Tpc = 156.5$~(1.5)~MeV~\cite{Bazavov:2018mes}
and $\Tpc = 158.0$~(0.6)~MeV~\cite{Borsanyi:2020fev}. In the limit of vanishing
two light quark masses, keeping the strange quark mass fixed to 
its physical value, the phase transition temperature in the continuum limit
has been found to be $T_c^{n_f=(2+1)}=132^{+3}_{-6}$~MeV~\cite{Ding:2019prx}.
A value consistent with this number has recently been
obtained also in calculations with twisted mass Wilson
fermions in a $(2+1+1)$-flavor calculation \cite{Kotov:2021rah}.

In three-flavor QCD calculations the boundary for a first
order transition has been found at $T_c^{n_f=3} = 134(3)$~MeV and a pseudoscalar mass, $m_\pi^c\simeq 110$~MeV \cite{Kuramashi:2020meg}. Irrespective of whether below this mass indeed a first order transition exists or whether this only corresponds to pseudo-critical temperature, this value for $T_c^{n_f=3}$ will drop further when
approaching the chiral limit, leading to a chiral transition temperature below that of (2+1)-flavor QCD.
In general it is expected that the
phase transition temperature drops with increasing
$n_f$. In fact, the FRG analysis presented in \cite{Braun:2006jd} suggests that the chiral phase transition temperature in QCD with $n_f=3$ massless quarks is smaller by about 25~MeV compared to the 2-flavor case.

In the study presented here we want to further explore
the nature of the chiral phase transition in 
$3$-flavor QCD and, in particular, provide a 
first estimate of the chiral phase transition 
temperature based on calculations with the highly improved staggered quark (HISQ) action. These 
calculations extend earlier studies performed with the HISQ action~\cite{Bazavov:2017xul} on coarser lattices. 
We work here at $\Nt = 8$ compared to $\Nt =6$
used in \cite{Bazavov:2017xul}.

The paper is organized as follows. In the next section we introduce the chiral observables we will study to determine the chiral phase transition temperature in three-flavor QCD. Section III summarizes basic relations needed for our discussion of finite size scaling (FSS)
of chiral observables. In Sec. IV we summarize some results on the chiral transition in three-flavor QCD  obtained previously on coarser lattices and present a 
first comparison with data on the disconnected part of the chiral susceptibility obtained in our new study.
In Sec. V we finally present our results for the FSS analysis of the chiral order parameter and its susceptibility, from which we deduce the chiral phase
transition temperature on lattices with temporal extent $N_\tau=8$. We give our conclusions in Sec. VI.

\section{Simulation Parameters and Observables}
\label{sec:observables}

\subsection{Simulation parameters}
The results that we present here were obtained from lattice QCD simulations with three degenerate flavors. In our calculations
with staggered fermions
we use the HISQ action and a tree-level improved Symanzik gauge action. This framework is identical to that used previously in finite
temperature calculations for (2+1)-flavor QCD 
\cite{Bazavov:2011nk,Bazavov:2012jq,Bazavov:2014pvz}
as well as the three-flavor QCD calculations \cite{Bazavov:2017xul} performed on lattices with temporal extent $N_\tau =6$.

The temporal extent of our lattices was fixed at $\Nt = 8$ throughout while the spatial
extent was chosen to be one of $\Ns = 24$, 32 or 40. In order to control finite-volume effects in our calculations we 
typically performed calculations at a given quark mass value for two different values of $\Ns$. The larger one is chosen such that
$m_\pi L\ge 3$ in the region of the pseudo-critical temperatures.

We used the Bielefeld GPU code \cite{Altenkort:2021fqk} to generate around 10,000-50,000 hybrid Monte Carlo trajectories separated 
by 0.5 Time Units (TU) for $\Ns = 40$ and 1 TU for the smaller volumes. These data sets have been
generated in 10 independent streams that have
been decorrelated initially using about 200
trajectories. The rational hybrid Monte Carlo
algorithm~\cite{Clark:2004cp,Clark:2005sq} was used to generate the configurations and the step sizes were tuned so as to achieve acceptance rates of 60-80\%.
We saved gauge field configurations after every
5th time unit and performed calculations of 
various chiral observables on these configurations.

We generated data sets for different values of 
the quark masses at up to 17 values of the 
temperature in the range $110~{\rm MeV}\lsim T\lsim 
170~{\rm MeV}$. As a guidance for our 
choice of bare quark masses we used the line
of constant physics (LCP) determined in 
Ref.~\cite{Bazavov:2014pvz} for the case of (2+1)-flavor QCD. This LCP defines the value
of the strange quark mass, $\ms(\beta)$,  as a function of the gauge coupling $\beta$. It is tuned to its physical value by demanding 
the mass of the $\eta_{s\bar{s}}$ meson to stay constant on this LCP. We choose different sets of three
degenerate light quark masses, $m_q$, corresponding to $m_q= H\ \ms(\beta)$, with
$H=1/27,\ 1/40,\ 1/60$ and $1/80$. Further
details of our scale setting are discussed
in the next subsection.

\subsection{Scale setting}

In the continuum limit the relation between
lattice cut-off and gauge coupling $\beta=10/g^2$ is controlled by the universal, asymptotic 2-loop $\beta$-function of three-flavor QCD,
\begin{equation}
 a\Lambda_L\equiv f(\beta)=\left( \frac{10 b_0}{\beta} \right)^{-b_1/(2 b_0^2)} \exp(-\beta/(20 b_0)) \; ,
 \label{2loop}
\end{equation}
with  $b_0=9/(16 \pi^2)$ and $b_1=1/(4 \pi^4)$
and $\Lambda_L$ denoting the QCD $\Lambda$-parameter for the three-flavor lattice discretization scheme.
Eq.~\eqref{2loop} receives corrections at non-zero values 
of the lattice spacing that depend on the
observable used to set the scale. We use here
the non-perturbative 
$\beta$-functions $f_Ka(\beta)$ and $a/r_1(\beta)$ determined in (2+1)-flavor QCD for the kaon decay constant, $f_Ka$, as well as the parameter $r_1/a$ deduced 
from the slope of the heavy quark potential
\cite{Bazavov:2014pvz}.

In all our figures we use the $f_K$ scaling
function to introduce a temperature on 
lattice with the temporal extent $\Nt=8$,
\begin{equation}
    T= \frac{1}{a\Nt} = \frac{\fK}{\fK a\Nt}\; ,
\end{equation}
where we set the scale for the temperature by
using the value  $\fK = 156.1/\sqrt{2}$~MeV 
for the kaon decay constant as has been done 
also in the determination of the chiral phase 
transition temperature in (2+1)-flavor QCD
\cite{Ding:2019prx}. We
note that this value agrees within errors with the Flavor Lattice Averaging Group (FLAG) average
\cite{FlavourLatticeAveragingGroup:2019iem}.
This scale setting also is used in all our fits to data. In some cases we use fits based on the $r_1/a$ parametrization to obtain some idea about systematic errors in our results that are not yet continuum extrapolated. As a physical
value for $r_1$ we use $r_1=0.3106$~fm.

As mentioned in the previous subsection we choose the three degenerate quark masses 
in our simulations as fractions of the strange
quark mass used in the (2+1)-flavor QCD
calculations to define a LCP. For this we use the parametrization of $\ms(\beta)$ given in 
Appendix C of Ref.~\cite{Bazavov:2014pvz}.
Our choice of three-flavor quark mass,
$\ms/80\ \lsim\ m_q\ \lsim\ \ms/27$ then corresponds to a light pseudoscalar Goldstone mass in the 
range $80~{\rm MeV}\ \lsim\ m_\pi\ \lsim\ 140~{\rm MeV}$ in the continuum limit of (2+1)-flavor QCD.

\subsection{Chiral observables}
On each saved gauge field configuration, we calculated the chiral condensate, $\pbp$, and its susceptibility,
$\chit$, which are obtained from the free energy density of three-flavor QCD, $f(T,m_q)= - (T/V)\ln Z(T,V,m_q)$, as first and second derivative with respect to the quark mass $m_q$,
\begin{eqnarray}
\pbp &=& - \frac{1}{\Ns^3\Nt}\frac{\partial \ln Z}{\partial m_q} \; , \\
\chit &=& \frac{\partial \pbp }{\partial m_q} \; .
\end{eqnarray}
The chiral susceptibility is represented in terms of
disconnected, $\chid$, and connected, $\chic$
contributions, $\chit  = \chid + \chic$.
These chiral observables are given in terms of the inverse of the staggered fermion matrix, $D_q(m_q)$, and its higher powers, 

\begin{align}
\pbp &= \frac{n_f}{4 N_\tau N_\sigma^3}\, \big\langle\tr\Minv_q\big\rangle, \notag \\
\chid  &= \frac{1}{N_\tau N_\sigma^3}\left(\frac{n_f}{4}\right)^2\left(\big\langle\tr^2\Minv_q\big\rangle - \big\langle\tr\Minv_q\big\rangle^2\right), \notag \\
\chic  &=-\frac{n_f}{4N_\tau N_\sigma^3}\big\langle\tr D^{-2}_q\big\rangle \; .
\label{eq:definitions}
\end{align}
We evaluate these chiral observables using up to 100 Gaussian random vectors to evaluate the trace
of the inverse of $D_q$.
After calculating the traces on each configuration, the observables were calculated by dividing the total number of configurations into 10 bins and using the jackknife procedure.

\begin{figure}[!tb]
\includegraphics[width=0.45\textwidth]{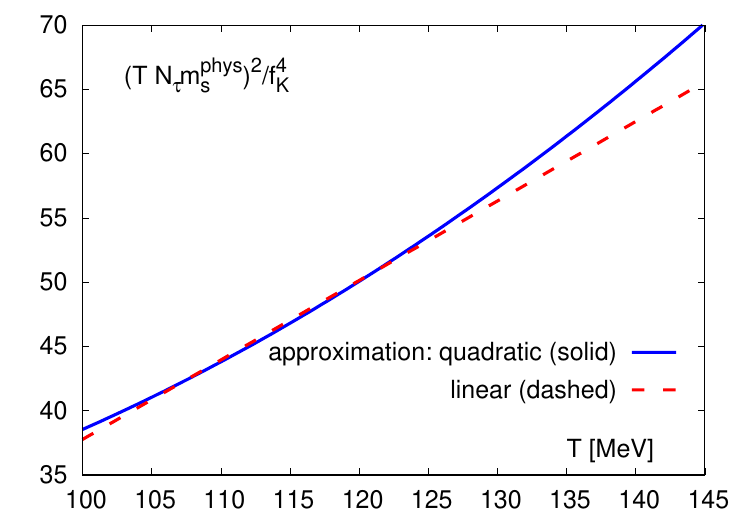}%
\caption{The prefactor of the UV-correction term 
for $N_\tau = 8$ and $c_{UV}=1$. Lines show quadratic and linear approximations of the parametrizations of $\ms (\beta)$ and $f_Ka(\beta)$ obtained in (2+1)-flavor QCD.}
\label{fig:cuv}
\end{figure}

In the continuum limit the chiral observables defined in Eq.~\ref{eq:definitions} require additive and multiplicative renormalization. The former arises 
from a ultra-violet divergent contribution to 
the chiral condensate, 
$\pbp=\pbp^{\text{ren}}+(c_{UV}/a^2) m_q$.
We use the strange quark mass $\ms(\beta)$ for
multiplicative renormalization of the chiral
observables to define the
order parameter, $M$, and its susceptibility, $\chi_{M}$ as,
\begin{eqnarray}
M &=& \ms\pbp^{\text{ren}}/\fK^4\; , \nonumber \\ 
\chi_{M} &=& \frac{\partial M}{\partial H} \; .
\end{eqnarray}
Here $\ms$ is our reference mass in three-flavor QCD, 
whose choice is motivated by the LCP determined
in (2+1)-flavor QCD calculations, and $H=m_q/\ms$.
This makes the observables dimensionless and RG-invariant up to logarithmic corrections.
As we do not have direct access to
the UV-divergent contribution, we instead evaluate
the unsubtracted observables
\begin{eqnarray}
M_b &=& \ms\pbp/\fK^4\; , \nonumber \\ 
\chi_{M_b} &=& \left( \ms\right)^2\chit/\fK^4 \; .
\end{eqnarray}
With this we have 
\begin{eqnarray}
    M&=& M_b - c_{UV}N_\tau^2 T^2 \frac{\left( \ms\right)^2}{\fK^4} H \; , \\
    \chi_{M} &=& \chi_{M_b} - c_{UV}N_\tau^2 T^2 \frac{\left(\ms\right)^2}{\fK^4} \; .
    \label{chiMbansatz}
\end{eqnarray}
Here the $1/a^2$ divergence manifests itself as $N_\tau^2$, since the continuum limit is obtained by
sending $N_\tau\to\infty$ at fixed $T$. For fixed, $N_\tau$, the UV-term picks up a quadratic 
temperature dependence and a logarithmic correction
arising from the anomalous dimension of the strange
quark mass term $\ms$. In fact, in a small 
temperature interval around the chiral phase 
transition temperature the temperature dependence 
of the UV-term is to a good approximation linear in $T$.
This is shown in Fig.~\ref{fig:cuv}.

Using the parametrizations of the strange quark mass
in lattice units, $\ms (\beta)$, and the kaon decay 
constant, $f_Ka(\beta)$, given in \cite{Bazavov:2014pvz} we evaluated the normalization 
factor multiplying the UV constant $c_{UV}$ and 
interpolated the result with a quadratic ansatz.
For this we find
\begin{equation}
    N_\tau^2 T^2 \frac{\left( \ms\right)^2}{\fK^4}
    =38.54 \left(1 + 1.23 \delta T+1.31 \left(\delta T\right)^2 \right) \; ,
\end{equation}
with $\delta T=(T-100)/100$.
This quadratic approximation is shown as a solid line in Fig.~\ref{fig:cuv}. Here we also give
the result of a linear fit in the temperature 
interval $[100:125~{\rm MeV}]$.
We note that this UV-term can be treated as part of the regular contributions in a (finite-size) scaling analysis. We will discuss this in more detail in the next section.

In order to eliminate the divergent UV contributions
explicitly in chiral observables we introduce the
difference
\begin{equation}
    M_\chi = M-H \chi_M \equiv M_b - H \chi_{M_b}\;\; .
\label{chiraldif}
\end{equation}
In fact, this observable can also be considered as
an order parameter for the chiral phase transition.
At high temperatures it vanishes, $M_\chi \sim H^3$,
and at low temperatures it equals the order parameter
$M$ at $H=0$ but receives different corrections at ${\cal O}(\sqrt{H})$.

\section{Scaling and finite-size scaling of chiral observables}
\label{sec:scaling_theory}

As there is increasing evidence that also three-flavor QCD will have a second order phase transition in the chiral limit, or at a rather
small quark mass, it is appropriate to analyze
thermodynamic observables in three-flavor QCD also
in terms of relevant scaling functions. 
As we are currently analyzing the chiral limit
at fixed values of the cut-off the universality class of 3-dimensional, $O(2)$ symmetric models 
would be appropriate, while the $Z(2)$ universality
class is of relevance, if a second order phase
transition occurs at a small value of the quark mass, $m_q^c$.

In the vicinity of a critical point, $(T_c,\mc)$, the thermodynamic free energy, $f(T,m_q)$, can be resolved into
singular and regular contributions, $f(T,m_q) = f_s(T,m_q) + f_r(T,m_q)$.
The temperature and quark mass dependence of $f_s(T,m_q)$ is expressed 
in terms of a
universal scaling function, $f_f(z)$, which is
characteristic for a particular universality class. Thus we have
\begin{equation}
f_s(T,m_q) = h_0 h^{1+1/\delta} f_f(z) \;\; ,
\;\;  z = t/h^{1/\beta\delta}\; ,
\end{equation}
with $t$ and $H$ being dimensionless variables constructed from the temperature $T$ and quark mass $m_q$,
\begin{equation}
t = \frac{1}{t_0}\frac{T-T_c}{T_c}\; , \;
h = \frac{1}{h_0}\frac{m_q-\mc}{\ms} \equiv \frac{H-H_c}{h_0}.
\end{equation}
The constants $\beta$ and $\delta$ are critical exponents of the 3D, $O(2)$ universality class for which we use \cite{Engels:2000xw}, 
\begin{equation}
    \beta = 0.3490 \;\; , \;\; \delta = 4.7798\; ,
\end{equation}
while $t_0$ and $h_0$ are
non-universal constants that are introduced to fix the overall normalization of the order parameter $M$~\cite{Ejiri:2009ac}.

For small quark masses, in the vicinity of the chiral transition temperature, $T_c$, the dominant contributions to the 
order parameter $M$ and its susceptibility $\chi_M$ 
arise from the singular part of the free energy.
Scaling relations for these observables are then
obtained by taking derivatives of $f_s(z)$ with respect to $H$. We have
\begin{equation}
M(T,m_q) = -\frac{\partial f_s}{\partial H}
= h^{1/\delta}f_G(z)\; ,
\label{eq:meos-i}
\end{equation}
and
\begin{equation}
\chi_M(T,m_q) = \frac{\partial M}{\partial H} = \frac{h^{1/\delta-1}}{h_0}f_\chi(z) \; ,
\label{eq:meos-ii}
\end{equation}
respectively. Here $f_G(z)$ and $f_\chi(z)$ are also universal functions of the scaling variable $z$, that
can be obtained from $f_f(z)$. Both these functions have been determined numerically using high-statistics Monte Carlo simulations
for the 3D, $O(2)$ and $O(4)$ universality classes~\cite{Engels:1999wf,Engels:2000xw,Engels:2003nq,Engels:2011km}. We will 
make use of the implicit parameterization provided for these functions for the $O(2)$ case
in Ref.~\cite{Engels:2001bq}. 

In the scaling regime the difference of order parameter and chiral susceptibility, introduced in Eq.~\ref{chiraldif}, is given in terms 
of these scaling functions
\begin{equation}
    M_\chi = h^{1/\delta} \left( f_G(z) -f_\chi(z) \right) \; .
\end{equation}

From Eqs.~\eqref{eq:meos-i} and \eqref{eq:meos-ii}, it is readily seen that
\begin{equation}
\frac{(H-H_c)\chi_M(0,h)}{M(0,h)} = \frac{f_\chi(0)}{f_G(0)} = \frac{1}{\delta}
\;\; ,
\label{eq:ratio}
\end{equation}
irrespective of the quark mass $m_q$. Curves for different quark masses will have a unique crossing point at 
$T=T_c$.

We also will analyze the related observable, obtained from the ratio of the non-subtracted chiral order parameter and its susceptibility,
$M_b/\chi_{M_b}$ for various $H$ at their corresponding pseudo-critical temperature \cite{Kaczmarek:2020err}.
The pseudo-critical temperatures defined by the maximum of the chiral susceptibility correspond to
a specific value of the scaling variable $z$, {\it i.e.} $z=z_p$,
\begin{equation}
\frac{M(T_{pc},h)}{\chi_M(T_{pc},h)} = (H-H_c) \frac{f_G(z_p)}{f_\chi(z_p)}
\label{eq:ratiozp}
\end{equation}
with $z_p=1.58(4)$ and $f_G(z_p)/f_\chi(z_p)=1.58$ and $1.51$ in the $O(2)$ and $Z(2)$ universality classes, respectively.
We note that the RHS of Eq.~\ref{eq:ratiozp} is solely determined by the universal contribution
which is an important aspect for the analysis of the nature of the chiral transition \cite{Kaczmarek:2020err}.

Note that Eqs.~\ref{eq:ratio} and \ref{eq:ratiozp} only hold provided the quark mass is sufficiently close to $\mc$ so that the
regular contributions coming from $f_r(T,m_q)$ can be ignored.

In addition to regular contributions, Eq.~\eqref{eq:ratio} also receives 
corrections when the system size $L$ is finite. Then the scaling
functions $f_G(z)$ and $f_\chi(z)$ in Eqs.~\eqref{eq:meos-i} and \eqref{eq:meos-ii}
must be replaced by the corresponding Finite-Size Scaling (FSS) functions
$f_{G,L}(z,z_L)$ and $f_{\chi,L}(z,z_L)$. Here $z_L = L_0/Lh^{\nu/\beta\delta}$ is a second scaling variable and $\nu=\beta (\delta+1)/d$ with $d=3$
is the critical exponent controlling the divergence of the correlation length at the critical point. $L_0$ is another
non-universal constant that can be fixed through a normalization condition, e.g. for 
the chiral condensate \cite{Engels:2014bra}. In the thermodynamic limit ($L\to\infty$) the finite size scaling functions,
$f_{G,L}(z,z_L)$ and $f_{\chi,L}(z,z_L)$, go over
to their infinite-volume equivalents 
$f_G(z)\equiv f_{G,L}(z,0)$ and $f_\chi(z)\equiv f_{\chi,L}(z,0)$.

The infinite and finite volume scaling functions have been determined for the 3D, $O(2)$ and $O(4)$ 
cases~\cite{Engels:2001bq,Engels:2014bra}.
In our FSS analysis we use a rational polynomial parametrization of the FSS functions \cite{MariusO2}  similar to what has been used also
for the analysis of FSS in $O(4)$ spin
models \cite{Engels:2014bra}

As we will see, the observed transition is a crossover for all the quark masses that we studied. According to
the standard picture of the phase diagram \cite{Brown:1990ev}, this transition should turn into a first order
phase transition if the quark mass is less than a certain value $m_q<\mc$. For $m_q=\mc$ then, the
transition will be second order belonging to the 3D, $Z(2)$ universality class. If on the other hand, the
expected first order region is absent, then the transition will be second order belonging to the 3D, $O(2)$ 
universality class in the three-flavor chiral limit for fixed $N_\tau$. Since we did not find any evidence of a
non-zero critical quark mass $\mc$ in our study, we will assume $\mc=0$ in the following and  use the $O(2)$
scaling functions for the rest of this work.

\section{Chiral order parameter and its susceptibility}
\label{sec:chiral_observables}

In the analysis of the chiral transition in
(2+1)-flavor QCD it has been found that in the
continuum limit the 
pseudo-critical temperature is about $T_{pc}\simeq 156.5$~MeV at physical values 
of the quark masses and drops in the 
limit of vanishing light quark masses to the
value of the chiral phase transition temperature of about $132$~MeV. On lattices with temporal
extent $\Nt=8$ the corresponding pseudo-critical
and chiral phase transition temperatures are larger by 5~MeV and 12~MeV, respectively. As we 
expect the pseudo-critical temperatures in three-flavor QCD to be smaller than those of the
(2+1)-flavor theory we explored in our 
calculations a range of temperatures
$100~{\rm MeV}\ \lsim\ T\lsim\ 170~{\rm MeV}$.

\begin{figure}[t]
\includegraphics[width=0.45\textwidth]{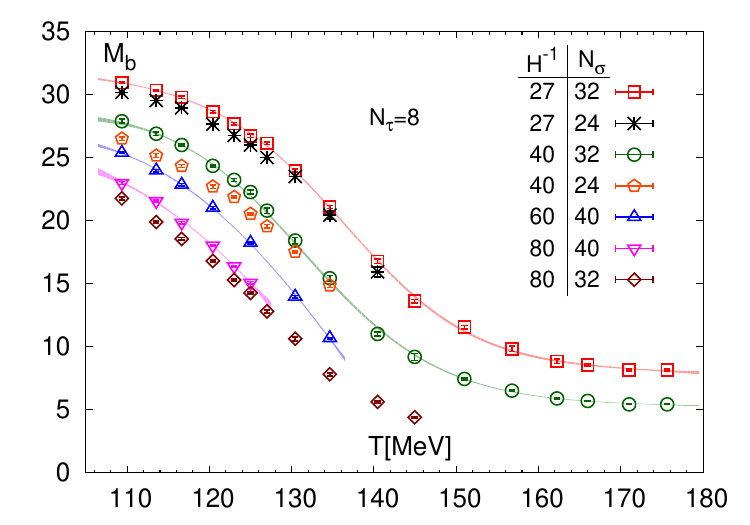}%
\hspace{0.05\textwidth}%
\includegraphics[width=0.45\textwidth]{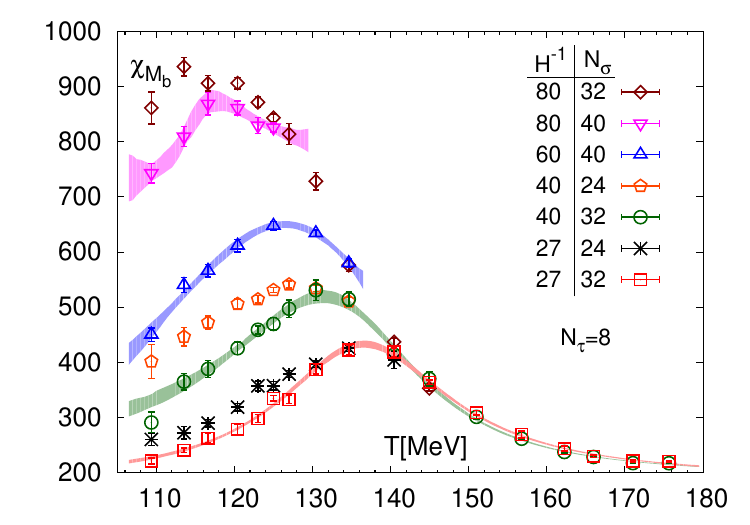}%
\caption{Results for the non-subtracted order parameter $M_b$ (top) and its susceptibility $\chi_{M_b}$ (bottom) as a function of
temperature for all quark masses and volumes. Bands show polynomial interpolations and are shown here
only for better visualization of the results on the largest lattice available at given quark mass.}
\label{fig:observables}
\end{figure}

\subsection{Chiral observables on \texorpdfstring{\boldmath$N_\tau=8$}{Nt=8} lattices}
We present our results for the non-subtracted chiral order parameter, $M_b$, and its susceptibility, $\chi_{M_b}$, calculated on lattices with temporal
extent $N_\tau=8$, in Fig.~\ref{fig:observables}. The order parameter $M_b$ varies rapidly
but smoothly, starting from a high value and decreasing to a low value over the temperature range considered here.
For three-flavor quark masses corresponding to $H=1/27$
we observe the most rapid change of $M_b$ at temperatures around $T\simeq (135-140)$~MeV. In this temperature range also the chiral susceptibility, $\chi_{M_b}$, has its maximum. This indicates that at this quark mass
value the pseudo-critical temperature in three-flavor QCD is shifted by about (20-25)~MeV
relative to that of (2+1)-flavor QCD. This trend also persists for the smaller quark masses examined by us.

Except for the case $H=1/60$ we show in Fig.~\ref{fig:observables} results for two different volumes. While the volume dependence is negligible in $M_b$ and $\chi_{M_b}$ slightly above
the corresponding pseudo-critical temperatures, 
evidence for a characteristic volume dependence 
is seen at smaller temperatures. In fact, contrary to what would be expected at or close to a second or first order phase transition, we see no increase in the peak height of $\chi_{M_b}$ with increasing volume. Instead we observe a slight decrease of the peak height of $\chi_{M_b}$ and a shift of the peak position towards larger temperatures as the volume is increased. At the same time the peak 
becomes more pronounced with increasing volume.
This behavior is reminiscent of the finite volume effects known from an analysis of finite
size scaling functions in 3-$d$, $O(N)$ symmetric spin models \cite{Engels:2001bq,Engels:2014bra}.

\begin{figure}[!tb]
\includegraphics[width=0.45\textwidth]{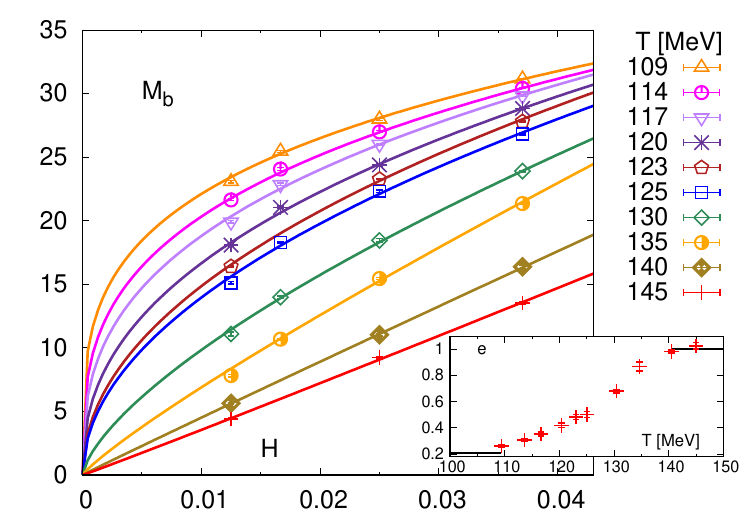}
\caption{The non-subtracted chiral order parameter $M_b$ as
function of $H$ at fixed temperature.
Lines show fits to the ansatz given in Eq.~\ref{eq:fit}. The inset shows results for the 
fit parameter $e$.
}
\label{fig:M_vs_m}
\end{figure}

At fixed quark mass we also see evidence in the data for a slight volume dependence of the 
non-subtracted chiral order parameter $M_b$.
This is clearly visible
in Fig.~\ref{fig:observables}~(top).
In Fig.~\ref{fig:M_vs_m} we show the chiral
order parameter as function of $H$
for several values of the temperature. As can be seen for $T\ge 140$~MeV
the order parameter depends linearly on the quark mass and extrapolates smoothly to zero for $H \rightarrow 0$. At lower temperatures the quark mass dependence of the
location of the maximum in $\chi_{M_b}$ suggests 
that in the chiral limit all our calculations correspond to a range of temperatures in the
chirally symmetric phase.

For a first analysis of the quark mass dependence 
of $M_b$
we thus fitted the data on the largest lattices available to an ansatz,
\begin{equation}
    M_b = A H^e \; .
    \label{eq:fit}
    \end{equation}
Results for the exponent $e$ are shown in the
inset of this figure. Within errors it is consistent with unity for $T\ge 140$~MeV
and decreases continuously with decreasing 
temperature. At the lowest temperature, $T\simeq 110$~MeV, we find $e\simeq 0.27$, 
which is larger but compatible with the 
exponent one expects to find at a 
critical point belonging to the 
3-dimensional $O(2)$ or $Z(2)$ universality classes, i.e. $e\equiv 1/\delta\simeq 0.21$. 

It is evident from Fig.~\ref{fig:observables}~(top) that finite 
volume effects are relevant at lower
temperatures and need to be treated carefully to arrive at results in the thermodynamic limit. We will analyze this volume 
dependence further in  Sec.~\ref{sec:scaling_theory}.

\subsection{Comparison with results from \texorpdfstring{\boldmath$N_\tau=6$}{Nt=6} lattices}

\begin{figure}[!tb]
\includegraphics[width=0.45\textwidth]{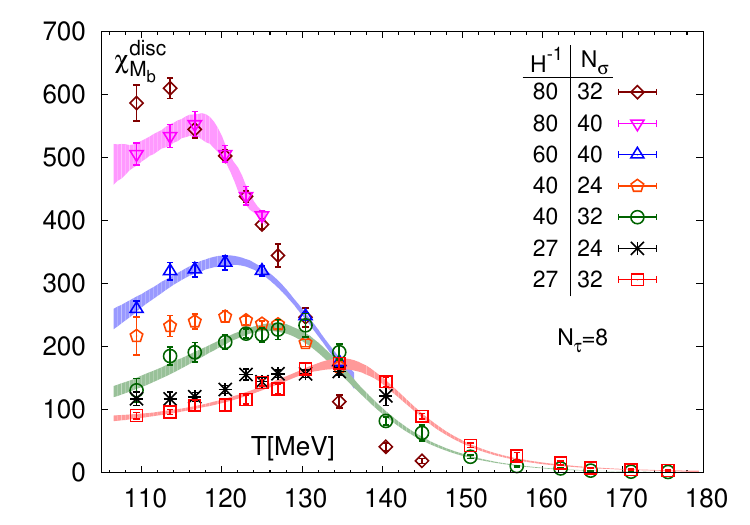}%
\caption{Results for the disconnected chiral susceptibility $\chi_{M_b}^\text{disc}$ as a function of
the temperature for all quark masses and volumes.}
\label{fig:chidis}
\end{figure}

Before going into a more detailed analysis of the volume dependence and universal scaling behavior 
of our results, obtained on 
lattices with temporal extent $N_\tau=8$, we 
want to compare with earlier results obtained
on coarser lattices with temporal extent $N_\tau=6$. In that case results exist only for 
the disconnected chiral susceptibility, $\chi_{M_b}^\text{disc}$, and critical couplings had been
extracted from the peak position of $\chi_{M_b}^\text{disc}$.
For better comparison with our current results 
we thus also show this susceptibility obtained by us on lattices with temporal extent $N_\tau=8$, in Fig.~\ref{fig:chidis}. In Tab.~\ref{tab:Tpc6} and \ref{tab:Tpc8}
we give pseudo-critical temperatures obtained 
from the peak positions of 
$\chi_{M_b}^\text{disc}$ for $N_\tau=6$ and $8$. 

In both
cases we used the temperature scale determined 
from the parametrization\footnote{We note
that the relevant range of couplings in $n_f=3$
calculations is smaller than those for 
$n_f=2+1$. Differences in temperature scales
extracted from different physical observables
thus are more pronounced. Comparing results
obtained from the parametrization of $a/r_1$ to
that of $f_Ka$ we
find that the former scale gives temperatures,
that are systematically larger by about 10~MeV for 
$N_\tau=6$ and 5~MeV for $N_\tau=8$. Of course,
in the continuum limit both ways of scale
setting will lead to a unique temperature scale
\cite{Bollweg:2021vqf}.}
of $f_Ka$.

The calculations for $N_\tau=6$~\cite{Bazavov:2017xul} have been
performed at various temperature values, keeping 
the quark mass fixed in units of the lattice spacing while in our current calculations on
$N_\tau=8$ lattices the quark mass is varied
along a line of constant physics by keeping 
the ratio $H$ fixed. Using the parametrization $\ms(\beta)$ we obtained the
corresponding ratio $H$ at the pseudo-critical
coupling $\beta_c(m_q)$ also for the $N_\tau=6$
data sets given in \cite{Bazavov:2017xul}. 
The resulting ratios $H^{-1}$ are given in
Tab.~\ref{tab:Tpc6}.
\begin{table}[t]
\begin{tabular}{|c|c|c|c|}
\hline
\multicolumn{4}{|c|}{ $N_\tau=6$} \\
\hline
$H^{-1}$ & $m_q$ & $N_\sigma$ & $T_{pc}^\text{disc}$ [MeV] \\
\hline
15.0 & 0.0075 & 16 & 138.2(4.4) \\
32.8 & 0.00375 & 24 & 129.8(1.1) \\
50.8 & 0.0025 & 24 & 126.9(2.7) \\
70.0 & 0.001875 & 24 & 124.2(1.3) \\
110.3 & 0.00125 & 24 & 120.2(1.3) \\
151.2 & 0.0009375 & 24 & 118.2(0.9) \\
\hline
\end{tabular}
\caption{\label{tab:Tpc6} Results for the pseudo-critical temperature
obtained from the location of peaks in 
$\chi_{M_b}^\text{disc}$ for $N_\tau=6$. They are obtained from
Ref.~\cite{Bazavov:2017xul} as discussed in the text.
}
\end{table}

\begin{figure}[t]
\includegraphics[width=0.45\textwidth]{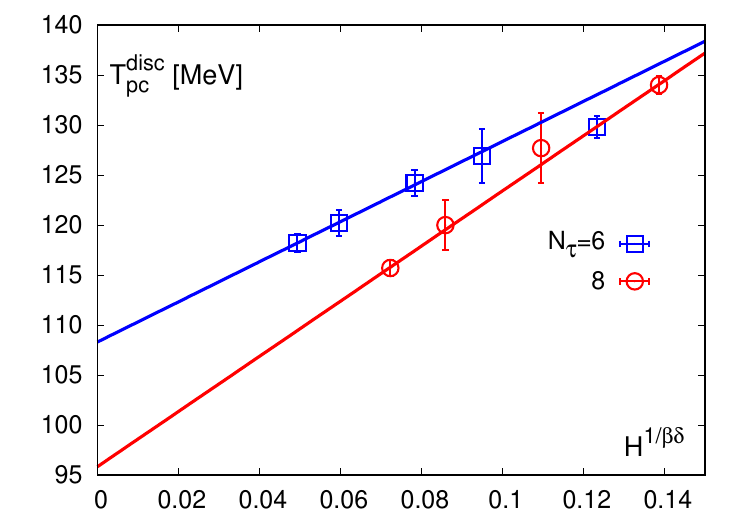}%
\caption{Pseudo-critical temperatures obtained
from the location of maxima in the disconnected
chiral susceptibility on the largest lattices available for various values of the quark masses
as given in Tab.~\ref{tab:Tpc6} and \ref{tab:Tpc8}
The straight line fits shown here are done for $H\le 1/40$ and $H\le 1/27$
for $N_{\tau}=6$ and $N_{\tau}=8$, respectively.
}
\label{fig:Tpc}
\end{figure}

\begin{table}[t]
\begin{tabular}{|c|c|c|c|}
\hline
\multicolumn{4}{|c|}{ $N_\tau=8$} \\
\hline
$H^{-1}$ & $m_q$ & $N_\sigma$ & $T_{pc}^\text{disc}$ [MeV]\\
\hline
27 & 0.00309 & 32 & 134.0(0.9)  \\
40 & 0.00220 & 32 & 127.7(3.5)  \\
60 & 0.00159 & 40 & 120.0(2.5)  \\
80 & 0.001237 & 40 & 115.7(0.8) \\
\hline
\end{tabular}
\caption{\label{tab:Tpc8} Results for pseudo-critical temperatures
obtained from the peak of the disconnected part of 
the chiral susceptibility
on the largest lattice used in simulations on lattices
with temporal extent $N_\tau=8$. 
}
\end{table}

The pseudo-critical temperatures, obtained for
$N_\tau=6$ and $8$ from maxima in the disconnected susceptibilities\footnote{Further details on the determination of these maxima are discussed in Section~\ref{sec:results}.B} are shown in
Fig.~\ref{fig:Tpc}. Here the data
are plotted versus $H^{1/\beta\delta}$, with 
$\beta\delta =1.67$ for the 3D, $O(2)$
universality class.
We fitted these data
using an ansatz inspired by the universal scaling of pseudo-critical temperatures in the vicinity of second order phase transitions \cite{Bazavov:2011nk},
\begin{equation}
T_{pc}^\text{disc}(m_q) = T_c+ a_c H^{1/\beta\delta} \left(
1 -b_c\ H^{1- 1/\delta  + 1/\beta\delta}  \right)
\; .
\label{Tc_zero}
\end{equation}
The term proportional to $b_c$ arises
from the leading regular contribution to the 
free energy, being proportional to $H$. However, as can be seen in
Fig.~\ref{fig:Tpc} our data are not sensitive
to this correction. In fact, they are 
well described by a straight line fit in terms of
$H^{1/\beta\delta}$. The fits shown in the figure for $b_c=0$ yield 
$T_{pc}^\text{disc}=108.3(4)$~MeV for $N_\tau=6$ and 95.8(7)~MeV for $N_\tau=8$, respectively. They 
both have a $\chi^2/$dof less than unity.

This result, as well as the quark mass 
dependence of the order parameter shown
in the previous subsections motivated a 
more detailed scaling analysis of the 
three-flavor results for the chiral order parameter and its susceptibility, which
we present in the following section.

\section{Finite Size Scaling Analysis of Chiral Observables}
\label{sec:results}
We want to improve here on our determination 
of the chiral transition temperature obtained
in the previous sections through an analysis
of disconnected part of the chiral susceptibility on the largest lattices available. 
We will analyze the finite size 
dependence of the unsubtracted chiral order parameter $M_b$
and its susceptibility $\chi_{M_b}$.

As can be seen in Fig.~\ref{fig:observables} 
$M_b(T,m_q,\Ns)$ and $\chi_{M_b}(T,m_q,\Ns)$ show a sizeable volume dependence 
for all the quark masses and for temperatures $T<\Tpc(m_q)$. Above the pseudo-critical temperature this volume dependence is significantly weaker. 
We define the pseudo-critical temperature,
$T_{pc}(m_q,\Ns)$, on lattices with spatial extent $\Ns$ as the location of the maximum of $\chi_{M_b}$ for a given quark mass and volume. We also define $\Tpc(m_q)$ as the infinite volume pseudo-critical temperature for a given quark mass. 

\subsection{FSS analysis of the chiral condensate
and its susceptibility}
\begin{figure}[t]
\includegraphics[width=0.48\textwidth]{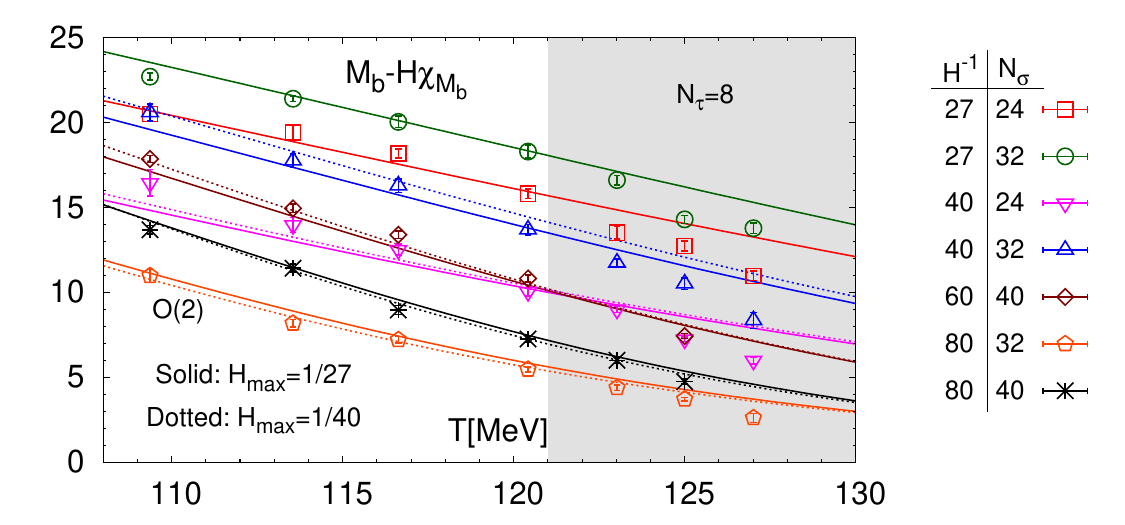}
\caption{Finite Size Scaling fits to the difference of
the chiral order parameter and its susceptibility,
$M_\chi(T,m_q,\Ns)$, with (solid lines) and without (dotted lines) the 
largest quark mass ratio $H=1/27$. The gray region has been left out of the fits. The fit parameters are summarized in Tab.~\ref{tab:pbpminusHtimessuscfitfss}.}
\label{fig:Mchi}
\end{figure}

We start with an analysis of the difference of the chiral order parameter and its susceptibility, introduced in Eq.~\ref{chiraldif}. As discussed there
this difference eliminates any dependence of the chiral observables linear in $H$ and thus also is independent of leading correction to scaling relations arising from
regular contributions to the chiral observables. In Fig. \ref{fig:Mchi}, we 
thus compare $M_\chi(T,m_q,\Ns)$ to the difference of
FSS functions,
\begin{equation}
    M_\chi(T,m_q,\Ns) = h^{1/\delta}\left( f_{G,L}(z,z_L) -f_{\chi,L}(z,z_L) \right) \; .
\end{equation}
Aside from the critical temperature in the chiral limit, $T_c$, a fit to this ansatz involves the three non-universal parameters, $h_0,z_0=h_0^{1/\beta\delta}/t_0,\ z_{L,0}=L_0h_0^{\nu/\beta\delta}$, which determine
the overall amplitude of $M_\chi$ and set the scale for the scaling variables $z$ and $z_L$, respectively.
It is apparent from the quark mass dependence of the maxima of $\chi_{M_b}$ and 
the estimate of $T_c$ from the chiral extrapolations in a finite volume, given in the previous sections, that the chiral phase transition will be located at a temperature 
below $110$~MeV. We thus perform fits only for
temperatures close of the chiral transition 
temperature, {\it i.e.} for temperatures below $121$~MeV.
The results of such fits in different fit ranges for
the quark masses, $H\in [0,H_\text{max}]$ are given
in Tab.~\ref{tab:pbpminusHtimessuscfitfss} and shown in Fig. \ref{fig:Mchi} . As 
can be seen the fits yield a good $\chi^2/$dof when
leaving out the data sets for our largest quark mass
ratio $H=1/27$. However, even when including these data sets we obtain a reasonable fit result for 28 data points using only the 
three non-universal parameters of the 3D, $O(2)$ scaling functions.

\begin{table}[!thb]
    \centering
    \begin{tabular}{|c|c|c|c|c|c|c|}
        \hline
        $H_{\text{max}}$ & fit & $T_c$ [MeV] & $h_0^{-1/\delta}$ & $z_0$ & $z_{L,0}$ & $\chisq$  \\
        \hline\hline 
        1/27 & singular & 100(1) & 72(1) & 0.50(3) & 4.05(7) & 4.52 \\ \hline
        1/40 & singular & 97(1) & 83(3) & 0.47(2) & 4.15(6) & 2.44 \\ \hline
        1/60 & singular & 93(2) & 94(6) & 0.41(3) & 4.16(7) & 1.29 \\ \hline
        1/80 & singular & 103(10) & 62(24) & 0.6(2) & 4.1(1) & 1.59 \\ \hline
    \end{tabular}
    \caption{Non-universal parameters of the $O(2)$ scaling functions obtained from fits of the difference of order parameter and chiral susceptibility, $M_\chi$.}
    \label{tab:pbpminusHtimessuscfitfss}
\end{table}

A related observable is the ratio of the chiral 
susceptibility and the chiral order parameter which
has been used in \cite{Ding:2019prx}
to determine the chiral phase transition temperature in
(2+1)-flavor QCD, 
\begin{equation}
    \frac{H \chi_{M_b}(T,m_q,\Ns)}{M_b(T,m_q,\Ns)} = \frac{f_{\chi,L}(z,z_L)+H^{1-1/\delta} f_\text{reg}(T)}{f_{G,L}(z,z_L)+H^{1-1/\delta} f_\text{reg}(T)} \; .
    \label{fGfchiratio}
\end{equation}
Unlike the difference $M_\chi$ analyzed above this ratio 
is sensitive to regular contributions as well as 
the subtraction of a UV divergent term. It, 
however, eliminates the explicit dependence on the non-universal scale parameter $h_0$.
For the contributions of the regular
term we use a leading order Taylor expansion in the vicinity of the chiral transition temperature, $T_c$,
\begin{equation}
    f_\text{reg}(T) = a_0 + a_1 \frac{T - T_c}{T_c}
    + a_2 \left(\frac{T - T_c}{T_c}\right)^2 \; .
    \label{freg}
\end{equation}
In the temperature range analyzed by us this regular term, also takes care of the UV divergent
contribution to the order parameter, as is apparent from Fig.~\ref{fig:cuv}. We note that in this 
fit ansatz the non-universal parameter $h_0$ does
not explicitly appear as an independent fit parameter, but is
absorbed as an overall factor $h_0^{1/\delta}$ in the fit parameters $a_i$ of the regular term.

\begin{figure}[t]
\includegraphics[width=0.48\textwidth]{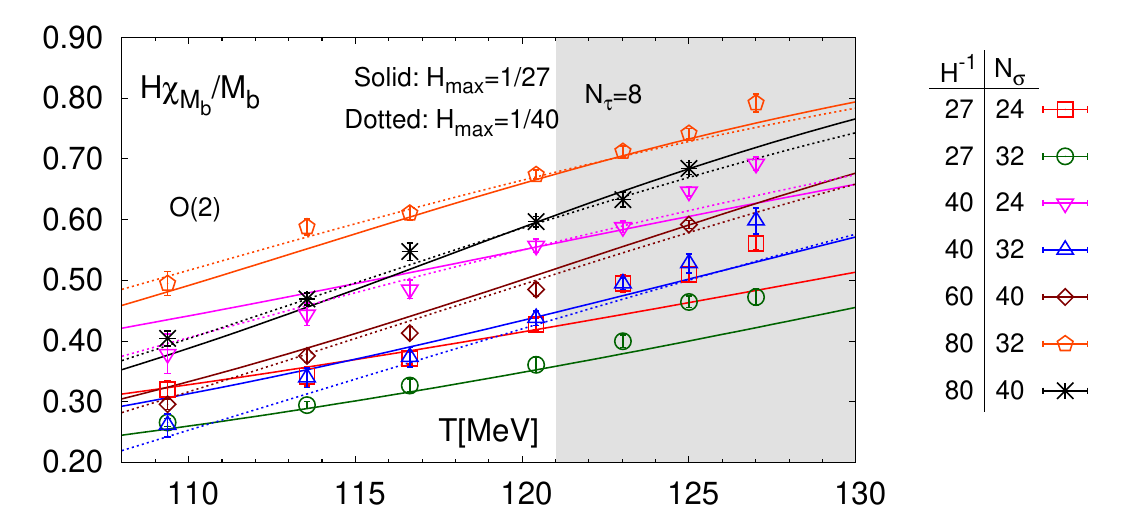}
\caption{Finite Size Scaling fits to the ratio of chiral susceptibility, $\chi_{M_b}(T,m_q,\Ns)$, and
chiral order parameter, $M_b(T,m_q,\Ns)$,  with (solid lines) and without (dotted lines) the largest quark mass ratio $H=1/27$. The gray region has been left out of the fits. Shown is a fit as given in Eq.~\ref{fGfchiratio} with a regular contribution linear
in temperature ($a_2=0$). The fit parameters are summarized in Tab.~\ref{tab:ratiofitfss}.}
\label{fig:chi}
\end{figure}

We perform a FSS analysis of the ratio given in
Eq.~\ref{fGfchiratio}.
We again performed fits in different fit intervals
$H\in [0,H_\text{max}]$ and a 
small temperature interval, $T/[{\rm MeV}]\in [0,121]$. In this temperature interval it suffices
to use a regular term, given by Eq.~\ref{freg}, with
$a_2=0$.
Results of these fits are summarized in the upper part of Tab.~\ref{tab:ratiofitfss} and 
are shown in Fig.~\ref{fig:chi} for the two cases $H_\text{max}=1/27$ and $1/40$, respectively.

\begin{table}[!thb]
    \centering
    \begin{tabular}{|c|c|c|c|c|c|c|}
        \hline
        $H_{\text{max}}$ & $T_c$ [MeV] & $z_0$ & $z_{L,0}$ & $a_0$ & $a_1$ & $\chisq$ \\ 
        \hline
                ~&\multicolumn{3}{c|}{singular part} &
        \multicolumn{2}{c|}{regular part} &~ \\
        \hline\hline 
        1/27  & 101(2) & 0.58(7) & 4.04(8)  & -0.4(4) & -0.4(3.2) & 2.97 \\ \hline
        1/40  & 93(4) & 0.37(7) & 4.15(6)   & -5(2) & -14(4) & 1.32 \\ \hline
        1/60  & 93(4) & 0.4(1) & 4.15(7)  & -4(2) & 6(8) & 0.77\\ \hline\hline
        1/27  & 101(1) & 0.57(4) & 4.04(8)  & -0.1(1) & -- & 2.85 \\ \hline
        1/40  & 100(1) & 0.56(4) & 4.17(7)   & -1.6(3) & -- & 1.89 \\ \hline    
        1/60  & 96(1) & 0.49(3) & 4.16(7)  & -3.1(4) & -- & 0.73\\ \hline  
        1/80 & 98(8) & 0.50(8) & 4.1(1)   & -0.8(5.5) & -- & 1.06 \\ \hline
        1/80 & 100(1) & 0.51(3) & 4.09(8)   & -- & -- & 0.85 \\ \hline        
    \end{tabular}
    \caption{Parameters of fits to the ratio $H\chi_{M_b}/M_b$ using the ansatz given in Eq.~\ref{fGfchiratio}. 
    The upper half shows fit results obtained
    with a regular term including terms linear in
    temperature, while in the lower half only results
    for fits with a constant term in the regular part
    are shown.
    For the smallest bound on the quark mass ratio, $H_\text{max}=1/80$, we also show the result from a fit the uses only the singular part of the fit ansatz.}
    \label{tab:ratiofitfss}
\end{table}

\begin{table*}[!thb]
    \centering
    \begin{tabular}{|c|c|c|c|c|c|c|c|}
        \hline
        $H_{\text{max}}$ & $T_c$ [MeV] & $z_0$ & $z_{L,0}$ & $a_0$ & $a_1$ & $a_2$ & $\chisq$ \\ 
        \hline
                ~&\multicolumn{3}{c|}{singular part} &
        \multicolumn{3}{c|}{regular part} &~ \\
        \hline\hline 
        1/27  & 101(2) & 0.55(6) & 4.09(6) & 2.0(7) & -39(8) & 139(20) & 3.45 \\ \hline
        1/40  & 91(4) & 0.33(6) & 4.15(5) & -0.5(2.1) & -27(15) & 81(26) & 1.51 \\ \hline
        1/60  & 91(4) & 0.33(7) & 4.14(5)  & -1.6(1.4) & -21(10) & 60(17) & 0.73 \\ \hline
    \end{tabular}
    \caption{ Parameters of fits to $H\chi_{M_b}/M_b$
    using a regular term quadratic in the reduced temperature $tt_0$. We give results for different
    fit ranges and cuts on the quark mass range:
    $T<135$ MeV for $H=1/27$, $T<130$ MeV for $H=1/40$, $T<126$ MeV for $H=1/60$.}
    \label{tab:ratiofitfsswithT2}
\end{table*}

We note that the fits based on differences of $M_b$ and $\chi_{M_b}$ and the fit of the ratio $H\chi_{M_b}/M_b$ are in excellent agreement with each
other, although only the latter is sensitive to regular contributions to the chiral observables.
The parameters of the singular part of these fits are consistent with each other within errors. The parameters of the regular term seem to be quite
sensitive to the upper bound ($H_\text{max}$) for the set of quark masses used in the fit. They are not well
determined in the small temperature interval used for
these fits. In fact, the $\chi^2/$dof changes only
little when only the leading temperature dependent 
term in the fit is used. The resulting parameters for
this fit are shown in the lower half of Tab.~\ref{tab:ratiofitfss}. 

For the case of $H_\text{max}=1/80$ the regular contribution vanishes within errors. We thus also give the result of a fit that only uses the singular term
in the fit ansatz. As can be seen this suffices to 
obtain a good fit at this small value of the quark mass and yields fit results for the non-universal 
parameters of the scaling functions that are in good
agreement with those obtained in a larger quark mass range by including contributions from a regular term.

We also performed fits that include a quadratic 
correction in the regular term. This allows us 
to enlarge the fit range for temperatures above the
transition temperature up to values that correspond to the pseudo-critical temperature for the largest quark mass contributing to the fit. This way we still obtain fits with $\chi^2/$dof
close to unity. Results of these fits are summarized 
in Tab.~\ref{tab:ratiofitfsswithT2}. We note that
these fits give large coefficients with opposite sign
for terms linear and quadratic in temperature. The 
reduced temperature factor $tt_0=(T-T_c)/T_c$ is about
1/3 at the upper end of the fit interval, $T\simeq 130$~MeV, which is comparable with the difference in
magnitude of the coefficients $a_1$ and $a_2$, {\it i.e.} 
contributions from linear and quadratic terms in the
regular part compensate each other to a large extent. We thus conclude that the parameters of the
singular part of our fits are well determined, while
the parametrization of the regular is not strongly
constrained.

\subsection{FSS analysis of the pseudo-critical temperature}
The chiral phase transition temperature $T_c$ can be determined from a finite size scaling analysis of the pseudo-critical temperature $\Tpcl(m_q,\Ns)$, which
can be obtained straightforwardly from our data for
$\chi_{M_b}(T,m_q,\Ns)$.

To determine $\Tpcl(m_q,\Ns)$ from the maxima of $\chi_{M_b}(T,m_q,\Ns)$, 100 bootstrap samples were constructed at the level of saved gauge configurations for each volume and quark mass, followed by fitting the $\chi_M$ data in the peak region to a quadratic ansatz for each such bootstrap sample. Starting with a minimum of three points, the fit interval was increased by adding one point at a time on either side of the peak to go up to a maximum of five points in all. In this way, we performed around 3-5 
fits for each bootstrap data set. The final value of $\Tpcl(m_q,\Ns)$ for a given quark mass and volume was obtained by taking
the mean of all the fits performed on the 100 bootstrap samples and error was the standard deviation of this distribution. The resulting estimates
for the finite-volume pseudo-critical temperatures for all available lattice sizes are given in Tab.~\ref{tab:allTpc8}.
Similar procedure was used to locate the peak position of $\chi_{M_b}^\text{disc}$ for the highest available volumes for each quark mass.

\begin{table}[t]
\begin{tabular}{|c|c||c|c||c|c|}
\hline
$H^{-1}$ & $m_q$ & $N_\sigma$ & $T_{pc}$ [MeV]& $\Ns$& $T_{pc}$ [MeV]\\
\hline
27 & 0.00309 & 24 & 136.0(1.2) & 32 & 136.8(1.9) \\
40 & 0.00220 & 24 & 127.9(1.9) & 32 & 131.9(1.2) \\
60 & 0.00159 & - & -- & 40 & 126.8(0.9) \\
80 & 0.001237 & 32 & 115.7(1.2) & 40 & 118.4(1.4) \\
\hline
\end{tabular}
\caption{\label{tab:allTpc8} Results for pseudo-critical temperatures
obtained from the maxima of the total chiral susceptibility on lattice with temporal extent 
$\Nt=8$ and different spatial lattice sizes.}
\end{table}

\begin{figure}[t]
\includegraphics[width=0.45\textwidth]{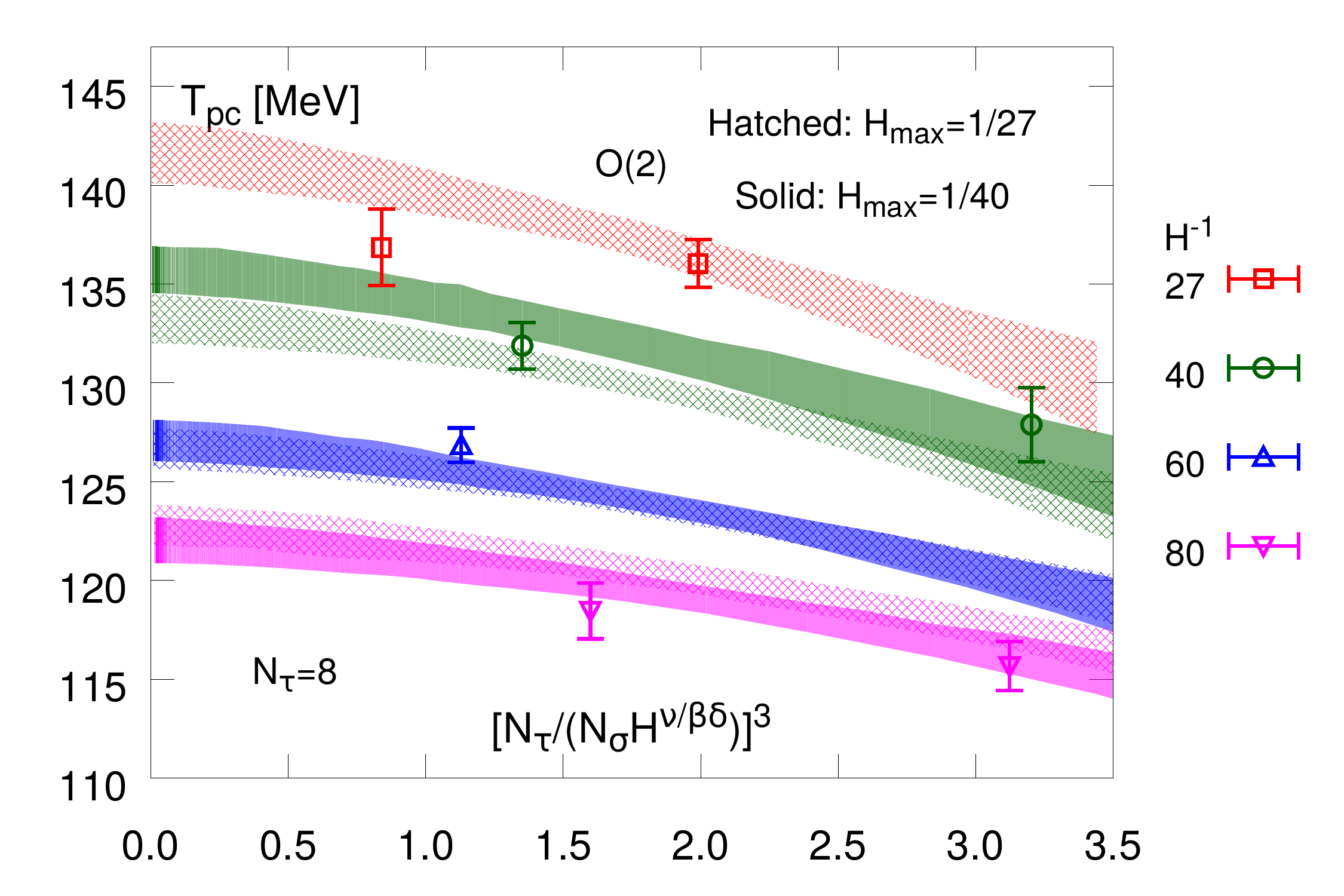}
\caption{Results for the peak location $\Tpcl(m_q,\Ns)$ plotted versus the finite volume scaling variable $(z_L/z_{L,0})^3 = (N_\tau/(\Ns H^{\nu/\beta\delta}))^3$. The data are plotted as points whereas the bands are the fit results for two different choices of the fit interval in $H$.}
\label{fig:peak_analysis}
\end{figure}

We present our results for $\Tpcl(m_q,\Ns)$ for all the quark masses and volumes in Fig.~\ref{fig:peak_analysis}.
We fitted these data using the FSS scaling ansatz
for $\Tpc$ deduced from the FSS ansatz for $\chi_{M_b}(T,m_q,\Ns)=\partial M_b/\partial H$ using Eq.~\ref{fGfchiratio},
\begin{equation}
    T_{pc}(m_q,\Ns) = T_c \left(1 +\frac{z_p(z_L)}{z_0} H^{1/\beta\delta} \right)  \; ,
    \label{Tpcfit}
\end{equation}
where $z_p(0)= z_p$ gives the location of the maximum of the infinite volume $O(2)$ scaling function $f_\chi(z)$ and $z_p(z_L)$ is a parametrization of the
finite volume dependence of this peak location \cite{MariusO2},
\begin{equation}
    z_p(z_L) = z_p \left( 1- 1.361(8)/z_L^{4.48(3)} \right) \; .
\end{equation}
This parametrization holds for $z_L \le 0.9$.
The resulting fits for the pseudo-critical temperatures
are also shown in Fig.~\ref{fig:peak_analysis}, again
for the cases with and without including the largest
quark mass, H=1/27, used in our calculations.
The chiral phase transition temperatures obtained from these fits are given in Tab.~\ref{tab:Tpcfssfit}.
The chiral phase transition temperature as well as the non-universal parameters
$z_0$ and $z_{L,0}$ obtained from the analysis of 
the maxima of $\chi_{M_b}$ are within errors in agreement with those deduced from our analysis of 
$M_\chi$ and the ratio $H \chi_{M_b}/M_b$.

\begin{table}[h]
    \centering
    \begin{tabular}{|c|c|c|c|c|}
        \hline
        $H_{\text{max}}$ & $T_c$ [MeV] & $z_0$ & $z_{L,0}$ & $\chisq$  \\ \hline
        1/27 &  102(2) & 0.57(5) & 3.8(2) & 3.58 \\ \hline
        1/40 & 95(3) & 0.41(5) & 3.7(2) & 2.68 \\ \hline
    \end{tabular}
    \caption{Results for the non-universal parameters 
    entering the fit to the pseudo-critical temperature, $\Tpc$, given in Eq.~\ref{Tpcfit}.}
    \label{tab:Tpcfssfit}
\end{table}

\begin{figure}[t]
\includegraphics[width=0.45\textwidth]{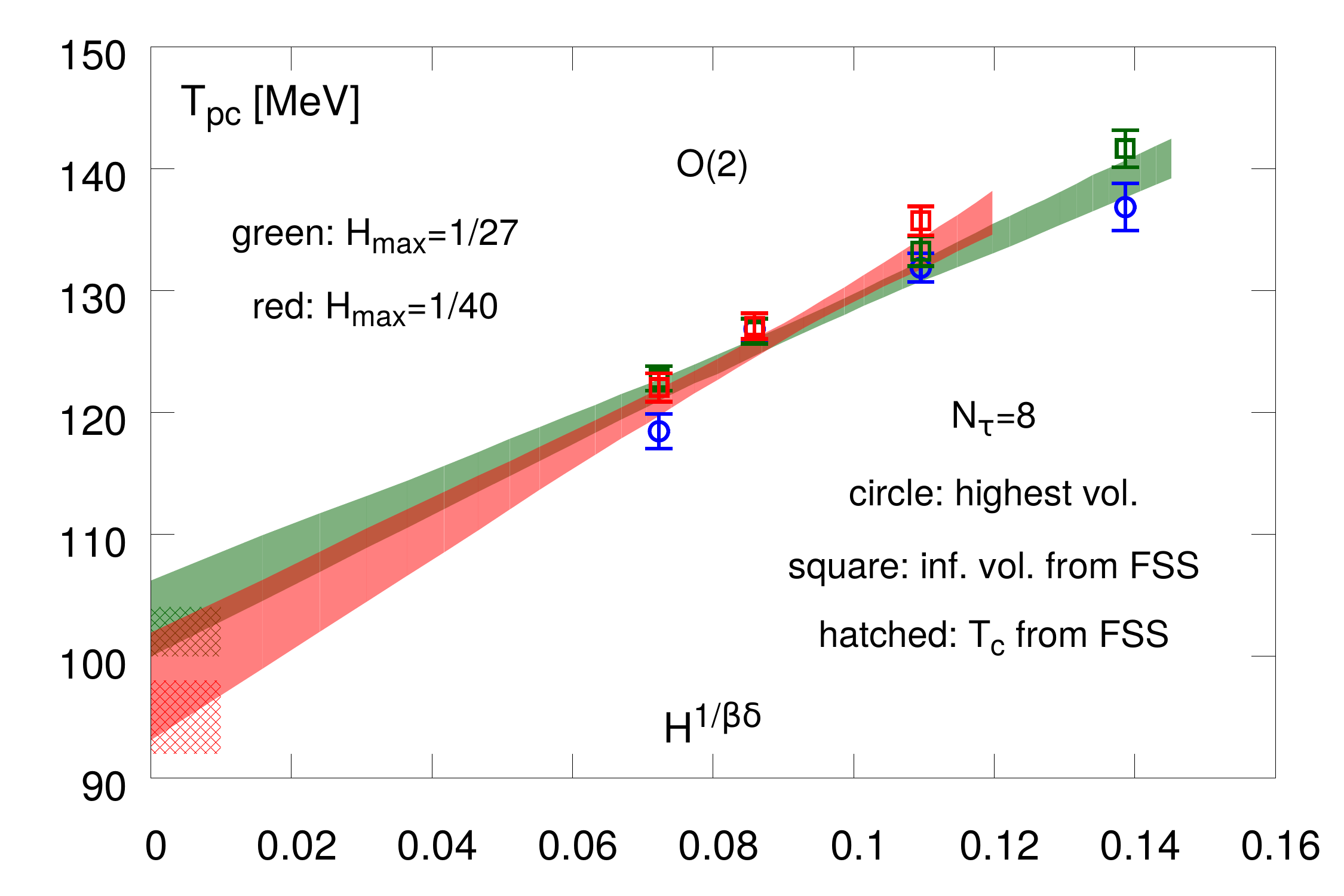}%
\caption{The pseudo-critical temperatures obtained
from the peak location $\Tpcl(m_q,\Ns)$ on the largest available volumes (circle), plotted versus $\Hbd$. Results at finite volume correspond to
$m_\pi L \simeq 4$ ($H^{-1}=27,\ 60$) and
$m_\pi L \simeq 3.3$ ($H^{-1}=40,\ 80$), respectively.
Also shown are results for the infinite volume
extrapolated pseudo-critical temperatures (squares) obtained from
the FSS analysis shown in Fig.~\ref{fig:peak_analysis}.
} 
\label{fig:TcpTc}
\end{figure}

In Fig.~\ref{fig:TcpTc} we show 
extrapolation of the pseudo-critical
temperatures to the chiral limit using $T_{pc}$
obtained on the largest volumes available for each mass. This gives $T_c=103(3)$~MeV and $97(4)$~MeV
with $H_\text{max}=1/27$ and $1/40$, respectively.
These numbers are in complete agreement with
the previously found numbers from the FSS fits.
In fact, the FSS fits shown in Fig.~\ref{fig:peak_analysis} suggest that the pseudo-critical temperatures, obtained on our largest 
lattices, differ from the infinite volume extrapolated results by less than 2 MeV.

\begin{figure}[t]
\includegraphics[width=0.45\textwidth]{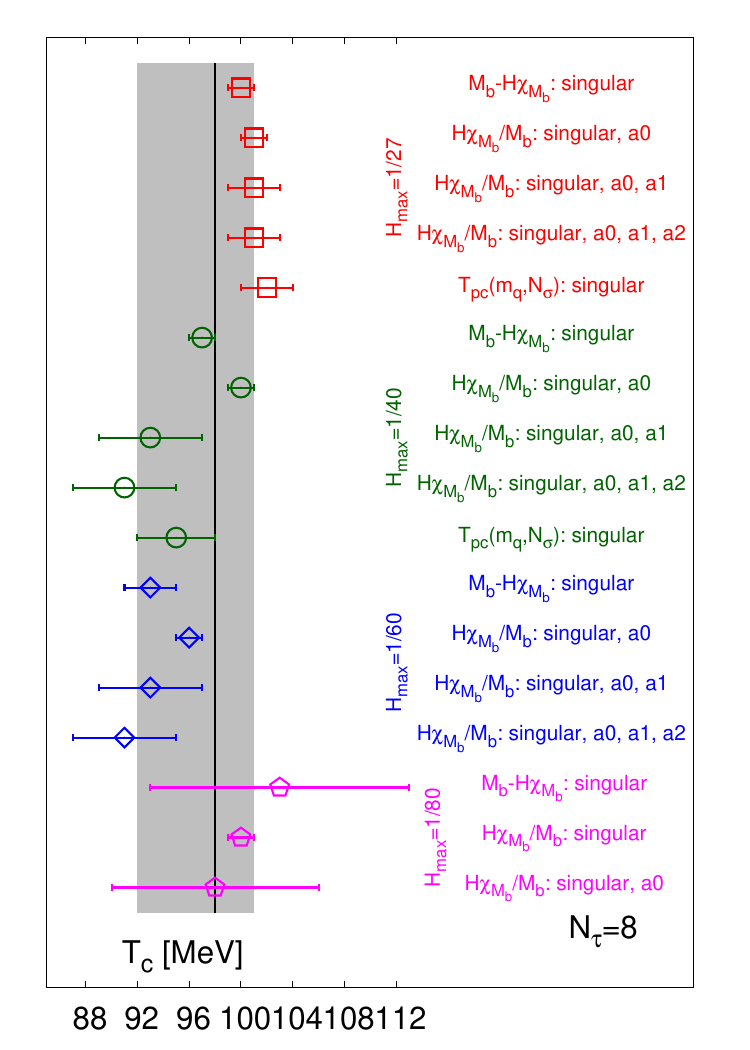}%
\caption{Summary of the fit results for $T_c$ obtained from various observables using different
fit ans\"atze and data sets as discussed in the text.
} 
\label{fig:finalTc}
\end{figure}

We took into account the systematic differences in
our fits resulting from changes of the fit range for $H$ as 
well as the fit ans\"atze that include or leave out contributions from
regular terms in the different observables we fitted. Averaging over all these fit results
for $T_c$  we obtain for the chiral phase transition temperature in three-flavor QCD on lattices 
with temporal extent $\Nt=8$,
\begin{equation}
    T_c = 98^{+3}_{-6}~{\rm MeV} \; .
    \label{eq.Tcval}
\end{equation}
This result is consistent within errors with all fit results for $\Tpc$ presented
in Tabs.~\ref{tab:pbpminusHtimessuscfitfss} to \ref{tab:Tpcfssfit} and shown in Fig.~\ref{fig:finalTc}.
Weighting with the Akaike information criterion or by the inverse of the squared error form each fit
yields estimates for $T_c$ which agree very well with the value quoted in Eq.\ \ref{eq.Tcval}.

\begin{figure}[t]
\vspace*{0.4cm}
\includegraphics[width=0.45\textwidth]{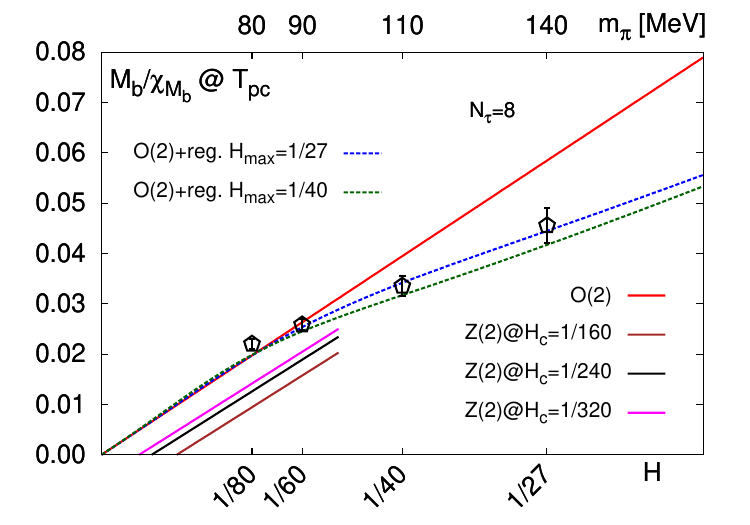}%
\caption{The ratio $M_b/\chi_{M_b}$ versus the quark mass ratio $H$ compared to scaling predictions from the $O(2)$ and $Z(2)$ universality classes. The dashed curves (blue and green) show the result of fits as discussed in the text.
} 
\label{fig:ratio}
\end{figure}

Our result for the chiral phase transition temperature, obtained from a scaling analysis 
for non-zero quark masses corresponding to
pseudoscalar Goldstone masses $m_\pi \gsim 80$~MeV,
relies on the occurrence of second order phase transition in the chiral limit. In support of this we 
show in Fig.~\ref{fig:ratio} the 
ratio $M_b/\chi_{M_b}$, evaluated at the position
of the maxima of $\chi_{M_b}$ for fixed $H$. 
The dashed curves, following the data, show the result of the fit
to $H\chi_{M_b}/M_b$ given in Eq.~\ref{fGfchiratio}, using a regular term up to quadratic order in $T$ and
evaluated at $T_{pc}$ obtained from Eq.~\ref{Tpcfit}, with $z_p(z_L)$ approximated by its infinite volume value $z_p(0)$.
This ratio is compared to straight lines with slopes given by the parameter ratio of 
scaling functions for the 3D, $O(2)$ universality
class (red) and three lines corresponding to the $Z(2)$ universality class, respectively. The latter is shown for three hypothetical critical masses $H_c$, below which a region of first order transitions might exist in $3$-flavor QCD.
This comparison puts stringent bounds on the possible
on a critical quark mass below which a first order
chiral phase transition may still occur. 
For quark mass ratios smaller than $H\simeq 1/60$
results for $H\chi_{M_b}/M_b$, evaluated at $T_{pc}(m_q,\Ns)$ are in good agreement with the parameter free universal $O(2)$ scaling function.
This leaves little room for a non-vanishing $H_c$
and a possible approach of the data for $M_b/\chi_{M_b}$ to the universal $Z(2)$ scaling 
function. Our results thus suggest a continuous
chiral phase transition in the 3D, $O(2)$
universality class at least at finite values of
the cut-off corresponding to $N_\tau=8$.

\section{Conclusions}
\label{sec:conclusions}

We have presented results on the chiral phase transition in three-flavor QCD. Our calculations have been
performed for finite values of the lattice spacing, corresponding to $N_\tau=8$.
For the range of  quark masses, corresponding in the continuum limit to light 
pseudoscalar Goldstone masses in the range $80~{\rm MeV} \le m_\pi \le 140~{\rm MeV}$ we find no direct
evidence for a conjectured first order phase transition. In the transition region we observe 
pseudo-critical behavior with a finite volume dependence that is consistent with the expected
FSS behavior in the 3D, $O(2)$ universality class. The parameter free comparison of the ratio
$M_b/\chi_{M_b}$ with 3D, $O(2)$ and $Z(2)$ scaling functions gives further support for a 
second order phase transition in the chiral limit of three-flavor QCD, at least on lattices with temporal extent $\Nt=8$.
For the chiral phase transition temperature at these non-vanishing values of the lattice spacing we find $T_c=98_{-6}^{+3}$~MeV. 

In (2+1)-flavor QCD the chiral 
phase transition at this value of the cut-off was about
10~MeV larger than the continuum limit extrapolated
phase transition temperature. Assuming that
cut-off effects are of similar magnitude also in three-flavor QCD, in the continuum limit the chiral phase transition
temperature in $3$-flavor QCD is expected to be below 90~MeV. This needs to be confirmed in future 
calculation. 

We also find that the pseudo-critical temperature at 
a pseudoscalar mass of about 110~MeV, which corresponds to $H\simeq 1/40$, is about (30-40)~MeV
larger than $T_c$, {\it i.e.} it is about 135~MeV as 
shown in Fig.~\ref{fig:peak_analysis}. This is indeed
consistent with the estimate of the transition temperature at this value of the pseudoscalar mass 
obtained with Wilson fermions \cite{Kuramashi:2020meg}. In this case, however, the transition temperature
is identified as the location of the end point of a region of first order phase transitions.
The difference between the two results obtained within the staggered and Wilson fermion discretization schemes, respectively, will need to be investigated further in the future.

The data corresponding to the plots in this work can be found in Ref. \cite{data_pub}.

\vspace{0.2cm}

\acknowledgements
This work was supported by  
the Deutsche Forschungsgemeinschaft (DFG, German Research Foundation) - Project number 315477589-TRR 211 and the grant 283286 of the European Union.
This research used awards of computer time made available through:
(i) the GPU-cluster at the Centre for High Energy Physics of the Indian Institute of Science, Bangalore, India;
(ii) the GPU-cluster at Bielefeld University, Germany;
(iii) a  PRACE grant at CINECA, Italy;
(iv) the Gauss Center at NIC-J\"ulich, Germany. 
We thank the Bielefeld HPC.NRW team for their support.

\appendix

\bibliography{refs}
\end{document}